\documentclass[twocolumn,showpacs,preprintnumbers,aps, prb,amsmath,amssymb]{revtex4}
\usepackage[dvips]{graphicx}
\begin{document}
{\bf Comment on ``Random-Field Spin Model beyond 1 Loop: 
A Mechanism for Decreasing the Lower Critical Dimension" \\
}

\begin{abstract}
A comment on the Letter by Le Doussal and Wiese, Phys. Rev. Lett. {\bf 96},197202
(2006).
\end{abstract}

%


In a recent letter [1], several interesting results 
in the random field O($N$) spin model near four dimensions
are obtained by a two-loop functional renormalization group. 
The existence of nonanalytic fixed points with a linear cusp
is one of them, which is shown by a large $N$ analysis
with the renormalization group.
It is argued that several of these fixed points
are once or twice unstable and they
yield a long crossover or the metastability of the system 
within a glassy region.
In this comment, 
however, we indicate that 
these nonanalytic fixed points are unphysical.
Here, we present our understanding of the phase transition 
in this model
for large $N$.

In $d=4+\epsilon > 4$ dimensions, the random field spin model has 
a random anisotropy function $R(z)$ as an effective interaction. As discussed
in the letter [1],
the fixed point 
condition in the large $N$ limit is expressed as
\begin{equation}
 -\tilde{R}(z) + 2 \tilde{R}'(1) \tilde{R}(z)
-\tilde{R}'(1) \tilde{R}'(z)z +\frac{1}{2}\tilde{R}'(z)^2=0,
\end{equation}
up to two-loop order 
in terms of 
a rescaled random anisotropy function $\tilde{R}(z)=
\lim_{N \rightarrow \infty} NR(z)/\epsilon$.
Denoting $y(z)=\tilde{R}'(z)$ and $y_0=y(1)$, a family
of the fixed points can be
parameterized by an integer $n \geq 2$, such that
\begin{equation}
y_0=\frac{n}{n-1},  \ \ \ \ \ \ z= y-(y_0-1)\left(\frac{y}{y_0}\right)^n.
\end{equation}
An even $n$  corresponds to the random field fixed point, and
a stability analysis
based on the eigenvalue problem of 
the scaling operator indicates that 
it has $n$ relevant modes [2].
An odd  $n$ corresponds to the random anisotropy fixed point, 
and it has $(n-1)/2$ relevant modes. 
In the random field systems, however, the correlation function exponents 
should satisfy
the Schwartz-Soffer inequality [3,4]
$\bar{\eta} \leq 2 \eta.$ 
This inequality is obtained by a simple
argument on the
correlation functions of the random field systems. 
The formulae for the correlation exponents [4]
\begin{equation}
\eta = R'(1), \ \ \ \ \ \bar{\eta}= (N-1) R'(1)- \epsilon
\label{formulae}
\end{equation}
and the Schwartz-Soffer inequality imply
$
R'(1) \leq \frac{\epsilon}{N-3},
$
which is valid for any finite $N$. 
Thus, $y_0 \leq 1$ in the large $N$ limit, 
if the corresponding fixed point is physical for finite $N$. 
Therefore the nonanalytic fixed points with 
$
y_0 = \frac{n}{n-1} > 1,$  for finite $n = 2, 3, 4, \cdots
$ 
cannot give any useful information for large but finite $N$. 
As long as the model has any small random field,
this upper bound gives useful restriction on the model.
If one discusses a model with a special constraint $R(-z)=R(z)$ to 
forbid the random field, one has no upper bound. 
Here, we do not discuss such a
random anisotropy model. 
Therefore, the only physical fixed points 
are the trivial fixed point $y(z)=0$,
the dimensional reduction fixed point
$y(z) = 1$ and
the second rank random anisotropy fixed point
$y(z) = z$, in the large $N$ limit. 

The stability analysis 
indicates that $y(z)=0$ is fully stable, $y(z)=1$
is once unstable and $y(z)=z$ is fully unstable [2]. 
For large, but finite $N$,
the fixed points with sufficiently large $n$ may satisfy the 
Schwartz-Soffer inequality. However,
they are unstable as well as the fixed point $y(z)=z$.  
On the other hand at the fixed point $y(z)=1$, 
infinitely many relevant modes recognized 
as serious instabilities during these two decades
are understood as a weak singularity of a nonanalytic fixed point
by Tarjus and Tisser (TT fixed point)[5]. This 
fixed  point is the unique once unstable fixed point given by
$R_{\rm TT}'(z) = R_{\rm DR}'(z) + a(1-z)^{\alpha}+ ....$,
where $\alpha= \frac{N}{2}-\frac{9}{2} +\cdots $ and $R_{\rm DR}(z)$
is 
analytic at $z=1$ and $R_{\rm DR}'(1)=\frac{\epsilon}{N-2}$.
The finiteness of $R_{\rm TT}(z)$ at $z=-1$ should fix the coefficient $a$.
Although neither a simple large $N$ expansion nor critical exponents 
can distinguish the TT fixed point from the dimensional reduction,
our stability analysis works to specify the TT fixed point.

For a sufficiently weak randomness, the renormalization group flow
is absorbed into the fully stable fixed point $R(z)=0$
universally. This fixed point characterizes the ferromagnetic phase.
The universal properties of the phase transition between the
ferromagnetic and disordered phases are classified into the following 
two cases.
In the first case for
$
N \geq 18-\frac{49}{5}\epsilon,
$
the flow goes into the unique once unstable TT fixed
point at the critical strength of the random field.
The dimensional reduction is observed in the critical exponents
$\eta = \bar{\eta}=\frac{\epsilon}{N-2}$.
The unique relevant mode 
is a finite eigenfunction 
with the eigenvalue 
$\epsilon+\frac{\epsilon^2}{N}+\frac{2\epsilon^2}{N^2} + \cdots$, which
confirms another prediction
$\frac{1}{\nu}= \epsilon + \frac{\epsilon^2}{N-2}+{\rm O}(\epsilon^3)$ 
by the dimensional reduction.
The amplification of this relevant mode leads to the disordered phase. 
  In the second case for 
$
N < 18-\frac{49}{5}\epsilon, 
$
the TT fixed point disappears.
It is believed that
the phase transition is
controlled by a nonanalytic fixed point with a linear cusp. 
The critical 
exponents $\eta$ and $\bar{\eta}$ are shifted from 
the values of the dimensional reduction [4].

\noindent
Yoshinori Sakamoto, 
Laboratory of Physics,
Nihon University
274-8501 Japan\\
\noindent
Hisamitsu Mukaida, 
Department of Physics, Saitama Medical College, 
350-0496 Japan\\
\noindent
Chigak Itoi, 
Department of Physics,
Nihon University, 
101-8308 Japan

\noindent
PACSnumbers: 75.10.Hk, 75.10.Nr, 05.50.+q, 64.60.Fr \\

\noindent
[1] P. Le Doussal and K. J. Wiese, Phys. Rev. Lett. {\bf 96}, 197202 (2006). \\
\noindent
[2] Y. Sakamoto, H. Mukaida and C. Itoi, 
Phys. Rev. B {\bf{72}}, 144405 (2005); 
{\it ibid} {\bf 74}, 064402, (2006). \\
\noindent
[3] M. Schwartz and A. Soffer, Phys. Rev. Lett. {\bf{55}}, 2499 (1985). \\
\noindent
[4] D. E. Feldman, Phys. Rev. B {\bf 61}, 382 (2000); 
Phys. Rev. Lett. {\bf{88}}, 177202 (2002).\\
\noindent
[5] Tarjus and Tisser, Phys. Rev.Lett.{\bf 93}, 267008 (2004).

\end{document}